\documentclass[twoside]{article}

\usepackage{fleqn,espcrc2, epsfig}

\def\fun#1#2{\lower3.6pt\vbox{\baselineskip0pt\lineskip.9pt
\ialign{$\mathsurround=0pt#1\hfil##\hfil$\crcr#2\crcr\sim\crcr}}}
\newcommand{\AmS}{{\protect\the\textfont2
  A\kern-.1667em\lower.5ex\hbox{M}\kern-.125emS}}
\newbox\mybox
\newcommand\fverb{\setbox\mybox=\hbox\bgroup\verb}
\newcommand\fverbdo{\egroup\medskip\noindent\fbox{\unhbox\mybox}\ }
\newcommand\fverbit{\egroup\item[\fbox{\unhbox\mybox}]}
\newcommand{\be}{\begin{equation}}
\newcommand{\ee}{\end{equation}}
\newcommand{\bc}{\begin{center}}
\newcommand{\ec}{\end{center}}
\newcommand{\ba}{\begin{eqnarray}}
\newcommand{\ea}{\end{eqnarray}}

\begin{document}

\title{The Kalman--Tran-D'Souza Model and the Semileptonic Decay Rates of Heavy
Baryons
\thanks{The invited talk presented by C.S.K. at the 4$^{th}$
International Conference on Hyperons, Charm and Beauty Hadrons,
Valencia, June 27--30, 2000. The work by P.Yu.K. and I.M.N. was supported in part by the
RFBR grants,
ref. No 00--02--16363 and 00-15--96786}}

\author{I. D'Souza$^{(a)}$, C. S. Kalman$^{(a)},$P. Yu Kulikov$^{(b){\ }}{and}$ I.
M. Narodetskii$^{(b){\ }}$ \\
$^{(a)}$Concordia University, Montreal, Canada \\
$^{(b)}$Institute of Theoretical and Experimental Physics, Moscow, Russia}
\begin{abstract}
We present an investigation of the inclusive semileptonic decay widths of
the heavy baryons \noindent $\Lambda _{Q}$ and $\Xi _{Q}$ $(Q=b,c)$ performed within a relativistic constituent quark model, formulated on the light-front.  
In a way conceptually similar to the deep-inelastic scattering case, the $%
H_{Q}$-baryon inclusive width is expressed as the integral of the free $Q$%
-quark partial width multiplied by a bound-state factor related to the $Q$%
-quark distribution function in the $H_{Q}$. The non-perturbative meson
structure is described through the quark-model wave functions, constructed
via the Hamiltonian light-front formalism using as input the
Kalman--Tran--D'Souza equal time wave functions. A link between
spectroscopic quark models and the $H_{Q}$ decay physics is obtained in this
way. It is shown that the bound-state effects and the Fermi motion of the $b$%
-quark remarkably reduce the decay rate with respect to the free-quark
result. Our predictions for the ${\rm BR}(\Lambda _{c}\to X_{s}\ell \nu
_{\ell })$ and ${\rm BR}(\Lambda _{b}\to X_{c}\ell \nu _{\ell })$ decays are
in good agreement with existing data.
\end{abstract}

\maketitle

\section{ Introduction}

\noindent The investigation of inclusive semileptonic $\Lambda _{b}$ decays
can provide relevant information on the Cabibbo-Kobayashi-Maskawa ($CKM$)
parameters $|V_{cb}|$ and $|V_{ub}|$ as well as on the internal
non-perturbative structure of the $\Lambda _{b}$ baryon. The first evidence
of semileptonic $\Lambda _{b}$ decays had been reported by ALEPH and OPAL
collaborations who had seen an excess of correlated $\Lambda _{s}\ell ^{-}$
pairs over $\Lambda _{s}^{+}$ pairs from $Z$ decays \cite{D92}, \cite{A92}.
More recently, the CDF collaboration \cite{A96} measured the $\Lambda _{b}$
lifetime using its semileptonic decay mode. It is therefore important to
study theoretical models for the semileptonic $\Lambda _{b}$ decays.

As far as the theoretical point of view is concerned, the $QCD$
-based operator product expansion ($OPE$) combined with the heavy quark
expansion is widely recognized as a consistent dynamical approach for
investigating inclusive heavy-flavour decays \cite{OPE}. Existing
theoretical predictions for inclusive weak decays are in reasonable
agreement, within the expected range of uncertainty, with the data on decays
of charmed and beauty particles. However, the result of the OPE approach for
the $\Lambda _{b}$ lifetime is puzzling because it predicts that $(\tau
_{\Lambda _{b}}/\tau _{B})_{OPE}=0.98+{\cal O}(1/m_{b}^{2})$, whereas the
experimental finding suggest a very much reduced fraction $(\tau _{\Lambda
_{b}}/\tau _{B})_{exp}=0.8\pm 0.04$. It is not clear whether the present
contradiction between the theory and the data on $\tau _{\Lambda _{b}}/\tau
_{B}$ is a temporary difficulty, or evidence for the fundamental flaws in
the OPE approach. In spite of great efforts of theoretical activity the $%
\Lambda _{b}$ lifetime remains significantly low which continues to spur
theoretical activity. In this respect, the use of phenomenological models,
like the constituent quark model, could be of great interest as a
complementary approach to the $OPE$ method.

What criteria should be used in determining an appropriate constituent quark
model for the baryons? There is now a vast amount of data on the spectrum of
baryons. Many quark model calculations of the spectrum attempt to account
for the masses, spin etc., without studying the decays. A serious drawback
to this approach is that it is difficult to assign predicted baryons to
experimentally known particles on the basis of spectrum alone. In this sense
this paper is a continuation of the work of Kalman and Tran \cite{KalTran} to
produce a complete picture of the spectrum and decays of the baryons. This
model is based upon the following principles: i) consistency with the model
of the mass calculation - A change of some feature of the mass calculation
in the decay model may compromise the purpose; ii) The decay calculation
should be mainly determined by the mass calculation. For instance, if there
are too many parameters in the decay model it is difficult to relate the
results obtained to the particles considered in the model, used in the mass
calculation. The decay calculations were based upon the quark pair creation
model of Le Yaouanc et. al. \cite{oliver} They studied the decay processes $%
N,\Delta \rightarrow N\pi $ and $\Lambda ,\Sigma \rightarrow N\overline{K}.$
The pair creation strength $\gamma $ is replaced by k$^{\gamma }$. Besides
the meson radius, these are the only parameters of the model. The results
are in good agreement with the data. To calculate decay widths in their
model, Kalman and Tran had to re-correct the originally calculated
parameters $\alpha _{\rho }$ and $\alpha _{\lambda }$ by approximately 0.05
GeV from the values determined from their spectrum calculation, in order to
make the decay widths match experimental data. The resulting change in $%
\alpha _{\rho }$ and $\alpha _{\lambda }$ causes a shift in the originally
calculated mass spectrum. It is these values that must be used if the decay
widths are to be correct.

At the time that the Kalman-Tran model was introduced, there was little data
available for the b and c sectors. The model cannot fit these sectors
properly without the modifications performed in the Kalman and D'Souza model 
\cite{KD}. The essential modification is that the coupling constant of the
interaction is made to be scale dependent in a manner that does not require
knowledge of the behaviour at low energy scales and does not introduce any
new parameter. The calculations of the inclusive semileptonic widths and the
corresponding branching ratios found here introduce no new parameters. It is
based solely upon the eigenfunctions for baryon states found by Kalman and
D'Souza \cite{KD}

In what follows we assume that, instead of QCD with its complicated dynamics
of infinite number of degrees of freedom in the light cloud, we consider a
constituent bound--state problem of a heavy quark interacting with a lighter
ones {\it via} a potential. Then, using the formalism of the light-front
(LF) relativistic quantum mechanics it is possible to encode all the
nonperturbative QCD effects in a LF quark model wave function $\psi
(x,p_{t}^{2})$ of a heavy hadron. The internal motion of a heavy $Q$--quark
inside the heavy flavor meson is described by the distribution function $%
|\psi (x,p_{t}^{2})|^{2}$, which represents the probability to find a heavy
quark carrying a LF fraction x=p$_{Q}^{+}$/P$_{H_{Q}}^{+}$ of the meson
momentum and a transverse relative momentum squared p$_{t}^{2}$={\bf p}$%
_{t}^{2}$. In Ref. \cite{KNST99} this formalism has been used to establish a
simple quantum mechanical relation between the inclusive semileptonic decay
rate of the heavy hadron and that of a free heavy quark. In this paper we
shall compute the preasymptotic effects for the inclusive semileptonic
decays of the heavy baryons H$_{Q}$ in the framework of the LF quark model,
which is a relativistic constituent quark model based on the LF formalism.

\section{Basic formulae}
The approach of \cite{KNST99} relies on the idea of duality in summing over
the final hadronic states. It has been assumed that the sum over all
possible charm final states $X_{Q^{\prime }}$ can be modelled by the decay
width of an on--shell $Q$ quark into an on--shell $Q^{\prime }$ quark folded
with the $Q$--quark distribution function $f_{H_{Q}}^{Q}(x,p_{t}^{2})=|%
\varphi _{H_{Q}}^{Q}(x,p_{t}^{2})|^{2}$ The latter represents the
probability to find a $Q$ quark carrying a LF fraction $x$ of the hadron
momentum and a transverse relative momentum squared $p_{t}^{2}$. For the
semileptonic rates the above mentioned relation takes the form
\begin{equation}
{\frac{d\Gamma _{SL}(H_{Q})}{dt}}={\frac{d\Gamma _{SL}^{Q}}{dt}}R_{H_{Q}}(t),
\label{1}
\end{equation}
where $d\Gamma _{SL}^{Q}/dt$ is the free quark differential decay rate, $%
t=q^{2}/m_{Q}^{2}$, $q$ being the 4--momentum of the $W$ boson, and $%
R_{H_{Q}}(t)$ incorporates the nonperturbative effects related to the Fermi
motion of the heavy quark inside the hadron. The expression for $d\Gamma
_{SL}^{Q}/dt$ for the case of non--vanishing lepton masses is given {\it e.g.%
} in \cite{KNST99}. $R_{H_{Q}}(t)$ in (\ref{1}) is obtained by integrating
the bound--state factor $\omega (t,s)$ over the allowed region of the
invariant hadronic mass $M_{X_{Q^{\prime }}}$ ($Q^{\prime }$ $=$ c or s) for
the underlying transitions $Q\rightarrow Q^{\prime }$ :
\begin{equation}
R_{H_{Q}}(t)=\int\limits_{s_{min}}^{s_{max}}ds\omega _{H_{Q}}(t,s),
\label{2}
\end{equation}
where $s=M_{X_{c}}^{2}/m_{b}^{2}$ and
\begin{eqnarray}
\omega _{H_{Q}}(t,s) &=&m_{Q}^{2}x_{_{0}}~{\frac{\pi m_{b}}{q^{+}}}~\frac{%
\left| {\bf q}\right| }{\left| {\bf \tilde{q}}\right| }  \nonumber \\
&&\bullet \int\limits_{x_{1}}^{\min \left[ 1,x_{2}\right] }dx|\varphi
_{H_{Q}}^{Q}(x,p_{t}^{*2})|^{2}.  \label{3}
\end{eqnarray}
In Eq. (\ref{3}) $x_{0}=m_{Q}/M_{H_{Q}}$, $x_{1,2}=x_{0}q^{+}/\tilde{q}^{\pm %
}$, 
where $q^{+}=q_{0}+|{\bf q|}$ is defined in the $H_{b}$ rest frame and $%
\tilde{q}^{\pm }=\tilde{q}_{0}\pm |{\bf \tilde{q}}|$ are defined in the $Q$
quark rest frame. In Eq. (\ref{2}) the region of integration is defined
through the condition $x_{1}\le min[1,x_{2}]$, {\it i.e.} $s_{min}=\rho
=(m_{Q^{\prime }}/m_{Q})^{2}$, $s_{max}=x_{0}^{-2}(1-x_{0}\sqrt{t})^{2}$.
For other details see \cite{KNST99}.

The structure of Eq. (\ref{1}) suggests that in the limit of heavy quarks
with infinite mass (i.e., $m_{Q}\to \infty $ and $m_{Q^{\prime }}\to \infty $%
) one has
\begin{equation}
\int dq_{0}~\omega (q^{2},q_{0})=1.  \label{4}
\end{equation}
which means that the total inclusive width of the hadron is the same as the
total inclusive width at the free quark level. The corrections to the
free-quark decay picture are mainly due to the difference between the quark
mass $m_{Q}$ and the hadron mass $M_{H_{Q}}$ as well as to the {\em %
primordial} motion of the $Q$-quark inside $H_{Q}$. These non-perturbative
corrections vanish in the heavy-quark limit $m_{Q}\to \infty $, but at
finite values of the $b$-quark mass a new parton description of inclusive
semileptonic decays, based on the constituent quark model, has been derived.
Note that the quark mass $m_{Q}$ appearing in\ref{3} is the{\it \ constituent%
} quark mass that differs from the {\it pole} quark mass $m_{Q,pole}$ that
enters the OPE result for the inclusive widths. As a consequence, Eq.\ref{1}
generally contains the correction to first order in 1/$m_{Q}$ expansion to
the free quark decay width. In principle, it is always possible to redefine
the constituent quark mass $m_{Q}$ in such a way that the 1/$m_{Q}$
correction is eliminated from the total semileptonic width in accordance
with the general statement of the OPE method. This is completely analogous
to the way used to eliminate 1/$m_{Q}$ corrections from the ACCMM model \cite
{BSUV94} and has been discussed within the considered approach in \cite
{KKNS99} for a particular case of the $B\to X_{s}\gamma $ decay width.

A priory, there is no connection between the equal--time (ET) wave function $%
w({\bf k}_{i})$ of a constituent quark model and LF wave function $\psi
(x_{i},p_{it}^{2})$. The former depends on the center--of--mass momenta $%
{\bf k}_{i}$, while the latter depends on the LF variables $x_{i}$ and ${\bf %
p}_{it}$. However, there is a simple operational connection between ET and
LF wave functions \cite{C92}. The idea is to find a mapping between the
variables of the wave functions that will turn a normalized solution of the
ET equation of motion into a normalized solution of the different looking LF
equation of motion. That will allows us to convert the ET wave function, and
all the labor behind it, into a usable LF wave function. This procedure
amounts to a series of reasonable (but naive) guesses about what the
solution of a relativistic theory involving confining interactions might
look like.

We convert from ET to LF momenta by leaving the transverse momenta
unchanged, ${\bf p_{it}}={\bf k_{it}}$, and letting $x_{i}=\frac{p_{i}^{+}}{%
P^{+}}$, where $P^{+}=\sum\limits_{i}p_{i}^{+}$. In the center--of--mass
frame $P^{+}=M_{0}$, where $M_{0}$ is the free mass operator, $%
M_{0}=\sum\limits_{i}\omega _{i}$ with $\omega _{i}=\sqrt{%
p_{it}^{2}+p_{iz}^{2}+m_{i}^{2}}$ and
\begin{equation}
p_{iz}=\frac{1}{2}(x_{i}-\frac{p_{t}^{2}+m_{i}^{2}}{x_{i}M_{0}^{2}})M_{0}
\label{5}
\end{equation}
The momentum fractions and the transverse momenta obey the conservation law

\begin{equation}
x_{1}+x_{2}+x_{3}=1,~~{\bf p}_{1t}+{\bf p}_{2t}+{\bf p}_{3t}=0  \label{6}
\end{equation}
Now we obtain the LF wave function from
\begin{equation}
\psi (x_{i},{\bf p}_{it})=\frac{\partial (k_{1z},k_{2z},k_{3z})}{\partial
(x_{1},x_{2},x_{3})}\cdot w({\bf k}_{1},{\bf k}_{2},{\bf k}_{3}),  \label{7}
\end{equation}
The straightforward calculation yields \cite{CS98}
\begin{equation}
\frac{\partial (k_{1z},k_{2z},k_{3z})}{\partial (x_{1},x_{2},x_{3})}=\frac{%
\omega _{1}\omega _{2}\omega _{3}}{x_{1}x_{2}x_{3}M_{0}}  \label{8}
\end{equation}
with

\begin{equation}
\frac{\omega _{i}}{x_{i}}=\frac{1}{2}\left( M_{0}+\frac{p_{it}^{2}+m_{i}^{2}%
}{x_{i}^{2}M_{0}}\right)  \label{9}
\end{equation}
It can be easily verified that
\begin{eqnarray}
&&\int\limits_{0}^{1}\left[ dx_{i}\right] \int \left[ dp_{it}\right] |\psi
(x_{i},{\bf p}_{it})|^{2}  \nonumber \\
&&\bullet \delta (\sum_{i}x_{i}-1)\delta (\sum_{i}{\bf p}_{i})=1
\end{eqnarray}
where

\begin{eqnarray}
\left[ dx_{i}\right] &=&dx_{1}dx_{2}dx_{3}\delta (\sum_{i}x_{i}-1),
\nonumber \\
\left[ dp_{it}\right] &=&d^{2}p_{1t}d^{2}p_{2t}td^{2}p_{3t}t\delta (\sum_{i}%
{\bf p}_{it})
\end{eqnarray}
provided
\begin{eqnarray}
&&\int \Pi _{i}d{\bf k}_{i}|w({\bf k}_{1},{\bf k}_{2},{\bf k}_{3})|^{2}
\nonumber \\
&&\bullet \delta ({\bf k}_{1}+{\bf k}_{2}+{\bf k}_{3})=1.
\end{eqnarray}
Wave functions made kinematically relativistic in this fashion are used to
calculate the form factors of heavy--to--heavy and heavy--to--light
exclusive transitions \cite{FF}.

\section{The Kalman--d'Souza model}

As was mentioned above, we consider the problem of a constituent bound state
formed by a heavy quark interacting with the lighter ones. Specifically, we
consider the Kalman--D'Souza model for heavy baryons \cite{KD}.

The constituent quark masses are
\begin{eqnarray}
m_{u} &=&m_{d}=0.23~{\rm GeV},~~m_{s}=0.605~{\rm GeV}~  \nonumber \\
~m_{c} &=&1.961~{\rm GeV},~~m_{b}=5.637~{\rm GeV}
\end{eqnarray}
We denote the light quarks (u,d, or s) by the labels 1,2 and the heavy quark
(c or b) by the label 3. For the cases of $\Lambda $ and $\Sigma $ baryons
where two of three quarks have equal masses we introduce the Jacobi relative
coordinates
\begin{equation}
\rho =\frac{1}{\sqrt{2}}({\bf r}_{1}-{\bf r}_{2}),~~~\lambda =\sqrt{\frac{2}{
3}}\left( \frac{m_{1}r_{1}+m_{2}r_{2}}{m_{1}+m_{2}}-r_{3}\right) 
\label{14}
\end{equation}
For the $\Lambda _{c}$ and $\Lambda _{b}$ baryons in the ground state, the
predominant configurations ($\approx 95\%$) of the heavy baryon wave
functions contain no excitations in $\rho $ and $\lambda $. Hence 
\begin{eqnarray} 
&&\Psi ({\bf \rho },{\bf \lambda })\approx \left( \frac{\alpha _{\rho }}{%
\sqrt{\pi }}\right) ^{3/2}\left( \frac{\alpha _{\lambda }}{\sqrt{\pi }}%
\right) ^{3/2}  \nonumber \\
&&\bullet \exp \left( -\frac{1}{2}\alpha _{\rho }^{2}{\bf \rho }^{2}\right)
\exp \left( -\frac{1}{2}\alpha _{\lambda }^{2}{\bf \lambda }^{2}\right),
\label{12}
\end{eqnarray}
where the factors $\alpha _{\gamma }$ ($\gamma =\rho ,\lambda $) are given
by
\begin{equation}
\alpha _{\gamma }=\alpha _{\gamma }^{(0)}\left( \frac{\mu _{\gamma }}{m_{1}}%
\right) ^{\frac{1}{4}},  \label{16}
\end{equation}
with
\begin{equation}
\mu _{\rho }=2\frac{m_{1}m_{2}}{m_{1}+m_{2}},~~~\mu _{\lambda }=\frac{%
3m_{3}\left( m_{1}+m_{2}\right) }{2\left( m_{1}+m_{2}+m_{3}\right) },
\label{17}
\end{equation}
and
\begin{equation}
\alpha _{\rho }^{(0)}=0.29~{\rm GeV},~~~\alpha _{\lambda }^{(0)}=0.32~{\rm %
GeV}.  \label{18}
\end{equation}

\noindent In Table 1, we compare the theoretical predictions of the
Kalman-D'Souza model with the experimental data.

\vspace{5mm}

{\bf Table 1}. The theoretical and experimental masses of the heavy baryons
and the parameters $x_{0}$, $\alpha _{\rho }$, and $\alpha _{\lambda }$ used
in the calculations. The theoretical masses are taken from \cite{KD}, while the experimental ones are taken from \cite{PDG}~(for $H_c$ and $\Lambda_b$) and from \cite{LAT} for $\Xi_b$\vspace{0.5cm}

\begin{center}
\begin{tabular}{|c|c|c|c|c|c|}
\hline\hline
$H_{Q}$ & $M_{th}$ & $M_{exp}$ & $x_{0}$ & $\alpha _{\rho }$ & $\alpha
_{\lambda }$ \\ \hline\hline
$\Lambda _{c}$ & 2.280 & $2.285\pm 0.0006$ & 0.86 & 0.29 & 0.40 \\ \hline
$\Sigma _{c}$ & 2.454 & $2.453\pm 0.0006$ & 0.80 & 0.29 & 0.40 \\ \hline
$\Xi _{c}$ & 2.461 & $2.466\pm 0.0014$ & 0.80 & 0.29 & 0.40 \\ \hline
$\Lambda _{b}$ & 5.640 & $5.624\pm 0.009$ & 1.00 & 0.29 & 0.41 \\ \hline
$\Sigma _{b}$ & 5.960 &  & 0.95 & 0.29 & 0.41 \\ \hline
$\Xi _{b}$ & 6.045 &  & 0.93 & 0.29 & 0.41 \\ \hline\hline
\end{tabular}
\end{center}
\vspace{5mm}

We now introduce the charge conjugate momenta
\begin{equation}
{\bf k}_{\rho }=\frac{1}{2}({\bf k}_{1}-{\bf k}_{2}),~{\bf k}_{\lambda }=%
\frac{m_{Q}({\bf k}_{1}+{\bf k}_{2})-2m_{q}{\bf k}_{3}}{2m_{q}+m_{Q}},
\label{19}
\end{equation}
then the momentum space wave function $\psi ({\bf k}_{\rho },{\bf k}%
_{\lambda })$ normalized according to (\ref{3}) is given by
\begin{eqnarray}
\psi ({\bf k}_{\rho },{\bf k}_{\lambda }) &=&\left( \frac{1}{\tilde{\alpha}%
_{\rho }\sqrt{\pi }}\right) ^{3/2}\left( \frac{1}{\tilde{\alpha}_{\lambda }%
\sqrt{\pi }}\right) ^{3/2}  \nonumber \\
&&\bullet \exp \left( -\frac{{\bf k}_{\rho }^{2}}{2\tilde{\alpha}_{\rho }^{2}%
}\right) \exp \left( -\frac{{\bf k}_{\lambda }^{2}}{2\tilde{\alpha}_{\lambda
}^{2}}\right)
\end{eqnarray}
with
\begin{equation}
\tilde{\alpha}_{\rho }=\frac{1}{\sqrt{2}}\alpha _{\rho },~~~~\tilde{\alpha}%
_{\lambda }=\sqrt{\frac{2}{3}}\alpha _{\lambda }  \label{21}
\end{equation}

\section{The heavy baryon light--front wave functions}

\noindent We denote x=x$_{3}$, y=x$_{1}$, {\bf p}$_{3}$= p$_{Q}$, then

\begin{eqnarray}
|\varphi (x,{\bf p}_{t})|^{2} =\frac{1}{8}\int\limits_{0}^{1-x}dy\int d%
{\bf p}_{1t}\hspace*{2cm}\nonumber \\
\bullet \left( M_{0}+\frac{\mbox{{\bf p}}_{Qt}^{2}+m_{Q}^{2}}{x^{2}M_{0}}%
\right)   
\left( M_{0}+\frac{\mbox{{\bf p}}_{1t}^{2}+m_{1}^{2}}{y^{2}M_{0}}%
\right)   \nonumber \\
\bullet \left( M_{0}+\frac{\left( \mbox{{\bf p}}_{Qt}+\mbox{{\bf p}}%
_{1t}\right) ^{2}+m_{2}^{2}}{\left( 1-x-y\right) ^{2}M_{0}}\right)
\left| \psi ({\bf k}_{\rho },{\bf k}_{\lambda })\right| ^{2}
\end{eqnarray}

\noindent where

\begin{center}
\begin{equation}
k_{\rho }^{2}=\left( {\bf p}_{1}+\frac{m_{1}}{m_{1}+m_{2}}{\bf p}_{Q}\right)
^{2}  \label{23}
\end{equation}
\begin{equation}
k_{\lambda }^{2}={\bf p}_{Q}^{2}={\bf p}_{t}^{2}+\frac{1}{4}\left( {\bf xM}%
_{0}+\frac{{\bf p}_{t}^{2}+m_{Q}^{2}}{{\bf xM}_{0}}\right) ^{2}  \label{24}
\end{equation}
\end{center}

\begin{figure}
\begin{center}
\epsfxsize=7.5cm
\epsfysize=7.5cm
\epsfbox{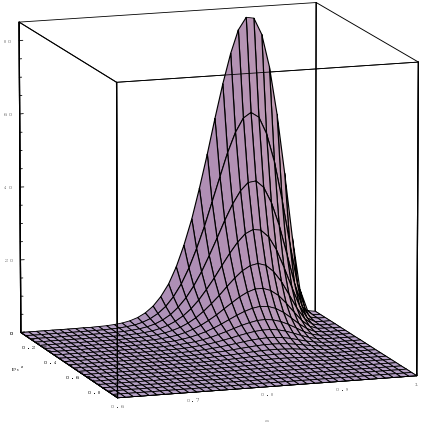} 
\end{center}
\vskip -1cm
\caption{Two--dimensional distribution function $f^b_{\Lambda_b}(x,p^2_t)$.}
\end{figure}

\begin{figure}
\begin{center}
\epsfxsize=7.5cm
\epsfysize=7.5cm
\epsfbox{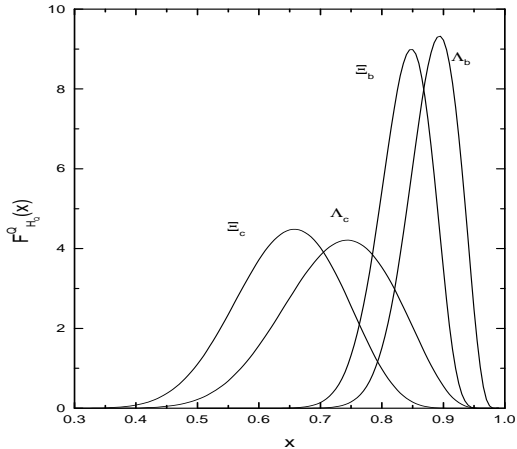} 
\end{center}
\vskip -1cm
\caption{One--dimensional distribution functions $F^Q_{H_Q}(x)$ for $\Lambda_c$ and $\Lambda_b$
baryons.}
\end{figure}

\begin{figure}
\begin{center}
\epsfxsize=7.5cm
\epsfysize=7.5cm
\epsfbox{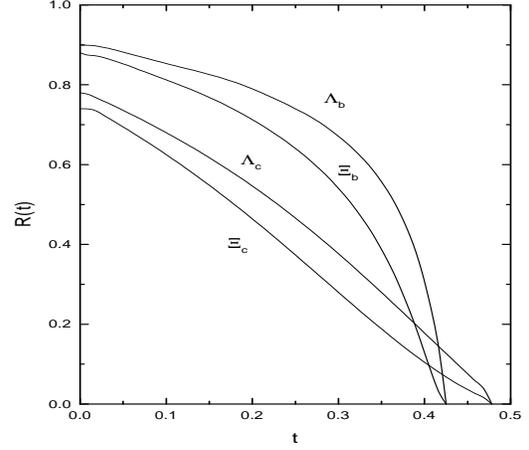} 
\end{center}
\vskip -1cm
\caption{The bound--state factor $R_{H_Q}$ in Eq. (\ref{2}).}
\end{figure}

In Fig.1 we show the two dimensional plot of the distribution function f$%
_{\lambda _{b}}^{b}$(x,p$_{t}^{2}$). In Fig.2 are shown the one dimensional
distribution functions F$_{H_{Q}}$(x), x=x$_{3}$. 

\begin{center}
\begin{eqnarray}
\label{df_Q}
F_{H_{Q}}(x) &=&\int\limits_{0}^{1}dx_{1}\int\limits_{0}^{1}dx_{2}\int
\prod\limits_{i=1}^{3}dp_{it}  \nonumber \\
&&\bullet |\psi _{H_{Q}}(x_{i},{\bf p}_{it})|^{2}  \nonumber \\
&&\bullet \delta (\sum {\bf p}_{it})\delta (\sum_{i}x_{i}-1).
\end{eqnarray}
\end{center}
\noindent for H=$\Lambda $,$\Xi $ and Q=b,c.

\noindent As illustrated in Fig.2 the
distributions f$_{\Xi _{c}}$ and f$_{\Xi _{b}}$are modified by the presence
of the s quark instead of u(d) one. In Fig. 3 we show the non--perturbative
factor R$_{H_{Q}}$(t) for $\Lambda _{b}$,$\Xi _{b},\Lambda _{c}$,and $\Xi
_{c}$baryons.\pagebreak

\section{The results}
We have evaluated Eqs.(\ref{1}-\ref{3}) in case of the inclusive semileptonic decays 
 $H_Q\to H_{Q'}
\ell\nu_{\ell}$ adopting the heavy quark distribution function $F_{H_Q}(x)$ from (\ref{df_Q}) with the Gaussian ans\"ats of the three-quark wave function (\ref{12}), 
that corresponds to the constituent quark model of ref. \cite{KD}. The parameters in 
(\ref{12}) have been taken from Table 1.
Our results for the  SL decay widths of $H_c$ and $H_b$ baryons and electron branching ratios 
$\Gamma_{SL}(H_Q)/\Gamma^{tot}(H_Q)$, where $\Gamma^{tot}(H_Q)=1/\tau(H_Q)$ is the total 
width of $H_Q$ baryon, are collected in Table 2. For the lifetimes we take the values 
from the recent PDG publication \cite{PDG} 
$\tau(\Lambda_c)=0.206~\pm~0.012~{\rm ps^{-1}}$,
$\tau(\Xi_c^+)=0.33^{~+0.06}_{~-0.04}~{\rm ps^{-1}}$,
$\tau(\Lambda_b)=1.229~\pm~0.08~{\rm ps^{-1}}$,
$\tau(\Xi_b)=1.39~\pm~0.3~{\rm ps^{-1}}$.

For the $c$--flavored baryons, there exists only one inclusive value for the semileptonic
branching ratio ${\rm BR}(\Lambda_c\to X_se\nu)= 4.5~\pm 1.7\%$, 
the corresponding exclusive semileptonic branching ${\rm BR}(\Lambda_c\to \Lambda e\nu)$ 
being $ 2.0~\pm 0.6\%$ \cite{PDG}.  
For $\Xi^+_c\to \Xi^0e\nu$ there exists only exclusive branching ratio 
$2.3^{+0.7}_{-0.9}\%$. 
For the $b$-flavored baryons, there exists  
the semileptonic branching ratio 
${\rm BR}(\Lambda_b\to \Lambda_c \ell\nu_{\ell}+anything)=7.9~\pm 1.9\%$ 
which is dominated by the charm baryons.  
In Table 2, in parentheses are also shown the results 
obtained   using OPE with 
the pole quark masses $m_c=1.6~{\rm GeV}$, $m_b=4.8~{\rm GeV}$ 
and $|V_{cb}|=0.039$ \cite{C97}.
We should mention a rather good agreement with the OPE predictions 
for the semileptonic rates in spite of the large 
difference of the constituent and pole quark masses that produces the large difference 
in $d\Gamma _{SL}^{Q}/dt\sim m_Q^4$. This difference 
is compensated by a factor $R(q^2)$ in Eq. (\ref{2}). 

\vspace{1.cm}

\noindent {\bf Table 2}. The predicted value of the inclusive
semileptonic widths (in units of ps$^{-1}$) and the corresponding
branching ratios in $\%$. The experimental data are from
\cite{PDG}. $|V_{cb}|=0.04$, $|V_{sc}|=1.00$. \vspace{0.5cm}

\begin{center}
\begin{tabular}{|c|c|c|c|}
\hline\hline
$ H_Q $ & $\Gamma (X_{Q'} e \nu_e) $ & $\Gamma(X_{Q'} \tau \nu_\tau) $ & 
$ BR(X_{Q'} e \nu_e)$  \\
\hline\hline
$\Lambda_c$ & 0.267~(0.233) & &5.50 \\
\hline
$\Xi_c$ & 0.236~(0.233) & &7.79  \\
\hline
$\Lambda_b$ & 0.086~(0.070) & 0.019 & 10.57 \\
\hline
$\Xi_b$ & 0.092~(0.070) & 0.021 & 12.79  \\
\hline
\hline
\end{tabular}
\end{center}

\section{Conclusions}

This is the most consistent quark model to-date describing baryons in the
sense that a single set of parameters is used for the whole spectra and then
subjecting all of their properties namely the masses, radii and their
wavefunction structures to a strict test by a decay model. The spectroscopy
is successfully calculated using far less parameters than any other model 
\cite{KD}. Decays of baryons are calculated \cite{KalTran}. Now in this paper
we have successfully computed the inclusive semileptonic decays of the heavy
baryons $H_{Q}$ by incorporating the LF formalism of \cite{KNST99}.

\end{document}